\documentclass[preprint,aps,12pt,showpacs,nofootinbib,tightenlines,amsmath,amssymb]{revtex4}
\usepackage{amsmath}
\usepackage{graphicx}
\usepackage{ulem}
\usepackage{color}
\begin{document}

\newcommand{\HC}{{\rm H.c}}
\newcommand{\TeV}{{\rm TeV}}
\newcommand{\GeV}{{\rm GeV}}
\newcommand{\CKM}{{\rm CKM}}

\title{$W^{\pm}H^{\mp}$ associated production at LHC in the general 2HDM with Spontaneous $CP$ Violation}

\author{Shou-Shan Bao$^{1,2}$, Yong Tang$^{1}$, Yue-Liang Wu$^{1}$ }
\affiliation{ $^{1}$ Kavli Institute for Theoretical Physics China (KITPC)
\\ Key Laboratory of Frontiers in Theoretical Physics, Institute of
Theoretical Physics, Chinese Academy of Science, Beijing,100190,
P.R.China \\
$^{2}$ Department of Physics, Shandong University, Jinan Shandong 250100, P.R.China}

\email{ylwu@itp.ac.cn}
\date{\today}

\begin{abstract}
Spontaneous $CP$ violation motivates the introduction of two Higgs
doublets in the electroweak theory. Such a simple extension of the
standard model has three neutral Higgs bosons and a pair charged Higgs, especially it leads to rich $CP$-violating
sources including the induced Kobayashi-Maskawa CP-violating phase, the mixing of the neutral Higgs bosons due to the $CP$-odd Higgs and the effective complex Yukawa couplings of the charged and neutral Higgs bosons. Within this model, we present the production of a charged Higgs boson in association with a $W$ boson at the LHC, and calculate in detail the cross section and the transverse momentum distribution of the associated $W$ boson.
\end{abstract}
\maketitle

\section{Introduction}
In the standard model(SM), the fermions and gauge bosons get masses
through Higgs mechanism with a single weak-isospin doublet Higgs
field. After the electroweak symmetry breaking, three Goldstone
modes are absorbed to build up the longitudinal W and Z gauge
bosons, and only one physical scalar called the SM Higgs boson is physical.
Since the exact breaking mechanism is not very clear and the Higgs haven't been detected yet, many extension models have been proposed.

One of the simplest extension of the SM is to add an
extra Higgs doublet motivated from Spontaneous $CP$
Violation(SCPV)\cite{TDLEE,SW,Liu:1987ng,YLW1,YLW2,Wu:1994ib}. It has been shown
that if one Higgs doublet is needed for the mass generation, then
an extra Higgs doublet is necessary for the Spontaneous $CP$ violation to explain the origin of $CP$ violation in SM. In such a model, the $CP$ violation is originated from a single relative phase of two vacuum expectation values, which not
only gives an explanation for the Kobayashi-Maskawa $CP$-violating
mechanism\cite{KM} in the SM, but also leads to a new
type of $CP$-violating source\cite{YLW1,YLW2}. Such a two Higgs
doublet model(2HDM) is also called Type III 2HDM to distinguish from the Type I and II 2HDM.

The common feature of the three types 2HDM is that after $SU(2)_L\otimes U(1)_Y$ gauge symmetry spontaneous breaking, there are three neutral Higgs and one pair charged Higgs. As shown in our previous work\cite{Bao:2009sa}, the neutral Higgs bosons decay to $b\bar{b}$ when their mass are light, which is difficult to be detected due to the strong background at the LHC. Therefore the charged Higgs boson ($H^\pm$) is of special interest, since there are no charged scalars in the SM and thus its discovery would constitute an indisputable proof of physics beyond the standard model. Thus the hunt for charged Higgs bosons will play a central role in the search for new physics at the LHC experiments.

Currently, the limit or constrain to the charged Higgs mass is not very strong and is also model-dependent. The best model-independent direct limit from the LEP experiments is $m_{H^\pm}>78.6\GeV$ (95\% CL)\cite{lep:2001ch} with assuming only the decays $H^+\to c\bar{s}$ and $H^+\to\tau\nu_\tau$. And as the charged Higgs will contribute to Flavor-Changing Neutral currents(FCNC) at one loop level, the indirect constrain can be extracted from B-meson decays. In Type II model, the constraint is $M_{H^\pm}\gtrsim350\GeV$ for $\tan\beta$ larger than 1, and even stronger for smaller $\tan\beta$\cite{typeII}. However, as the phases of the Yukawa couplings in Type III model are free, $m_{H^\pm}$ can be as low as 100\GeV\cite{BCK}. In this note we take it as free from 150GeV to 500GeV.

At the LHC, the interesting channels for the charged boson production are $gb \rightarrow H^{-}t$ for $m_{H^{\pm}} > m_{t}+m_{b}$ and $gg\rightarrow H^{-}t\bar{b}$ for $m_{H^{\pm}} \lesssim m_{t}-m_{b}$ \cite{Barnett,Bawa,Borzumati,Miller}.In these channels, the leptonic decay $H^+\to\tau^+\nu$ seems most promising for detecting light charged Higgs, while the hadronic decay $H^+\to t\bar{b}$ may be hopeful above threshold with efficient b-tagging \cite{Barger:1993th, Belyaev:2002eq,Assamagan:2004tt, Abazov:2008rn, Raychaudhuri:1995cc, Moretti:1996ra, Roy:1999xw, Guchait:2006jp}. Another interesting channel is to produce the $H^{\pm}$ in association with $W$ bosons, and the leptonic decays of the W-boson can serve as a trigger for the $H^\pm$ boson search. This channel can also cover the transition region search, $M_{H^\pm}\sim m_t$. The dominant channels for $W^{\mp}H^{\pm}$ production are $b\bar{b}\rightarrow W^{\pm}H^{\mp}$ at tree level and $gg\rightarrow W^{\pm}H^{\mp}$ at one-loop level. $W^{\mp}H^{\pm}$ production at hadron colliders in Type II 2HDM and MSSM has been extensively studied in \cite{Kniehl,Dicus,Moretti,Brein,Asakawa,Eriksson,Gao:2007wz}. The $CP$ violation effect is also explored at muon collider \cite{Akeroyd}. In this paper, we will study it in the Type-III 2HDM with Spontaneous $CP$ Violation(SCPV) \cite{YLW1,YLW2}. The discovery of relative light charged Higgs boson with $M_{H^\pm} < 350\GeV$ distinguishes from the type II 2HDM.

This paper is organized as follows. In the section II, we shall first give a brief introduction of the 2HDM with SCPV and some conventions. Then calculations are outlined in section III and numerical results are shown in section IV. Finally, we come to our conclusions.

\section{2HDM with SCPV}
 We begin with a brief introduction to the model by showing the spontaneous CP violation and its difference to Type I and Type II models. The two complex Higgs doublets are generally expressed as
\begin{equation}
\Phi_1=\left(\begin{array}{c} \phi_1^+\\ \phi_1^0
\end{array}\right), \, \, \, \Phi_2=\left(\begin{array}{c} \phi_2^+\\
\phi_2^0\end{array}\right),
\end{equation}
and the potential is
\begin{eqnarray} \label{eq:potential}
V(\Phi_{1}, \Phi_{2}) & = & -\mu_{1}^{2} \Phi_{1}^{\dagger}\Phi_{1}
-\mu_{2}^{2} \Phi_{2}^{\dagger}\Phi_{2} -(\mu_{12}^{2} \Phi_{1}^{\dagger}
\Phi_{2} + h.c.) \nonumber \\
& & + \lambda_{1}(\Phi_{1}^{\dagger}\Phi_{1})^{2}
+ \lambda_{2}(\Phi_{2}^{\dagger}\Phi_{2})^{2} +
\lambda_{3}(\Phi_{1}^{\dagger}\Phi_{1}\Phi_{2}^{\dagger}\Phi_{2})
+ \lambda_{4}(\Phi_{1}^{\dagger}\Phi_{2})(\Phi_{2}^{\dagger}\Phi_{1}) \\
& &  + \frac{1}{2} [\lambda_{5}(\Phi_{1}^{\dagger}\Phi_{2})^{2}  + h.c. ]
+ [(\lambda_{6}\Phi_{1}^{\dagger}\Phi_{1}
+ \lambda_{7} \Phi_{2}^{\dagger}\Phi_{2})(\Phi_{1}^{\dagger}\Phi_{2}) +
\HC].  \nonumber
\end{eqnarray}
With $\lambda_{5}$ non-zero and real, CP violation can arise from non-zero values of one or more of $\mu_{12}^{2}$, $\lambda_{6}$ or
$\lambda_{7}$. If these three (and $\lambda_{5}$) are all real, CP violation can  occur spontaneously \cite{TDLEE} when $\lambda_{5} > 0$, because of the relative phase between the vacuum expectation values. The most interesting case is that only the dimension-2 term $\mu_{12}^{2}$ is complex, which is known as a soft CP-violating phase. Then it can easily be demonstrated that once all other couplings in the Higgs potential are required to be positive real, the relative CP-violating phase of two vacuum expectation values is solely determined by the explicit soft CP-violating phase\cite{YLW2,Wu:1994ib} via the minimal conditions of Higgs potential. In this case, the CP violation remains originating from the spontaneous breaking of symmetry in vacuum, while it can avoid the so-called domain-wall problem.

The Yukawa interaction terms have the following general form
\begin{eqnarray}
-\mathcal{L}_{Y}
&=&\eta_{ij}^{(k)}\bar{\psi}^{i}_{L}\tilde{\Phi}_k U^{j}_{R}
  +\xi_{ij}^{(k)}\bar{\psi}^{i}_{L}\Phi_k D^{j}_{R}+\HC,\label{eq:oyukawa}
\end{eqnarray}
where $\psi^{i}_{L}=(U^{i}_{L},D^{i}_{L})^{\rm{T}}$, $\tilde{\Phi}_k=i\tau_2\Phi^{*}_k$, $\eta_{ij}^{(k)}$ and $\xi_{ij}^{(k)}$ are all real Yukawa coupling constants to keep the interactions $CP$-invariant. The above interactions may lead to Flavor-Changing Neutral Currents(FCNC) at the tree level through the neutral Higgs boson exchanges as the Yukawa matrices may not be diagonal. FCNC processes should be strongly suppressed based on the experimental observations. Usually, an {\it ad hoc} discrete symmetry \cite{discrete} is often imposed:
\begin{eqnarray}
  \Phi_1\to -\Phi_1 & \hspace{0.3cm}\textrm{and}\hspace{0.3cm} & \Phi_2 \to \Phi_2,\nonumber\\
  U_R   \to -U_R    & \hspace{0.3cm}\textrm{and}\hspace{0.3cm} & D_R    \to \mp D_R.
\end{eqnarray}
which correspond to Type-I and Type-II 2HDM relying on whether the up- and down-type quarks are coupled to the same or different Higgs doublet. Some interesting phenomena for various cases in such types of models without FCNC have been investigated in detail in \cite{barger1, barger2}. When the discrete symmetry is introduced in the potential Eq.(\ref{eq:potential}) , it leads to $\mu_{12}=0$ and $\lambda_6=\lambda_7=0$ and then no spontaneous $CP$-violation any more \cite{noscpv}. Since the FCNC is observed in experiments in the weak interactions though it is strongly suppressed, we shall abandon the discrete symmetry and consider the small off-diagonal Yukawa couplings. The naturalness for such small Yukawa couplings may be understood from the approximate global U(1) family symmetries \cite{YLW1,YLW2,FS,HW}. As if all the up-type quarks and also the down-type quarks have the same masses and no mixing, the theory has an U(3) family symmetry for three generation, while when all quarks have different masses but remain no mixing, the theory has the $U(1) \otimes U(1) \otimes U(1)$ family symmetries and the Cabibbo-Kobayashi-Maskawa quark-mixing matrix is a unit matrix, in this case both the direct
FCNC and induced FCNC are absent. In the real world, there are some FCNC processes observed, thus the U(1) family symmetries should be
broken down. As all the observed FCNC processes are strongly suppressed, the theory should possess approximate U(1) family symmetries with small off-diagonal mixing among the generations. In this sense, the approximate U(1) family symmetries are enough to ensure the naturalness of the observed smallness of FCNC.

As in the potential Eq.(\ref{eq:potential}), the neutral Higgs bosons will
get the vacuum expectation values as follows
\begin{equation}
\langle\phi_1^0\rangle=\frac{1}{\sqrt{2}} v_1 e^{i \delta_1}, \, \,
\, \langle\phi_2^0\rangle= \frac{1}{\sqrt{2}} v_2 e^{i \delta_2},
\end{equation}
where one of the phases can be rotated away due to the global U(1)
symmetry in the potential and Yukawa terms. Without losing generality, we may take $\delta_1=0$ and
$\delta_2=\delta$. It is then convenient to make a unitary
transformation
\begin{equation}\label{eq:transform}
\left(
\begin{array}{c}
H_1 \\ H_2
\end{array}
\right)
=U
\left(
\begin{array}{c}
\Phi_1 \\ \Phi_2
\end{array}
\right),
\hspace{0.15cm}\textrm{with} \hspace{0.15cm}
 U= \left(
\begin{array}{cc}
\cos\beta  & \sin\beta e^{-i\delta}\\
-\sin\beta & \cos\beta e^{-i\delta}
\end{array}
\right),
\end{equation}
where $\tan \beta =v_2/v_1$. After making the above transformation and redefining the $\phi^{0}_{i}$, we can re-express the Higgs doublets as follows:
\begin{equation}\label{eq:higgs}
H_1=\frac{1}{\sqrt{2}}\left( \begin{array}{c} 0\\ v+\phi^{0}_{1}
\end{array}\right)+\mathcal{G}, \, \, \, H_2=\frac{1}{\sqrt{2}}\left(
\begin{array}{c} \sqrt{2} H ^+\\ \phi^{0}_{2}+i \phi^{0}_{3}
\end{array}\right),
\end{equation}
with $v^2=v_1^2+v_2^2$ and $v\simeq 246\GeV$ which is the same as in
the standard model. Thus in this new basis, only the Higgs doublet $H_1$
gives masses to the gauge bosons, quarks and leptons. The Higgs field
$\mathcal{G}$ are the goldstone particles absorbed by the gauge
bosons, while $H^{\pm}$ are mass eigenstates of the charged scalar
Higgs. $\phi^{0}_{i}=(\phi_1^0, \phi_2^0, \phi_3^0)$ are the neutral Higgs
bosons in the electroweak eigenstates. In general, they are not the
same as the physics Higgs bosons $h_j=(h_1, h_2, h_3)$  in the mass eigenstates, but related via an
orthogonal SO(3) transformation
\begin{equation}
 \phi^{0}_{i}=O_{ij}h_{j} \hspace{0.2cm} \rm{with} \hspace{0.2cm} i,j=1,2,3,
\end{equation}
where $O_{ij} $ depends on the $\lambda_i$ and $\mu_i$ in the Higgs potential. When there is no mixing between $\phi_1^0, \phi_2^0$ and $\phi_3^0$, $h_1, h_2$ and $h_3$ are then corresponding to $h^0, H^0$(CP-even) and $A^0$(CP-odd) in the literature, respectively. For later discussion convenience, we will denote the mixing angle $\alpha$ of $h_1$ and $h_3$ by meaning that
\begin{equation}\label{eq:mixingalpha}
\left(
\begin{array}{c}
\phi_1^0 \\ \phi_2^0 \\ \phi_3^0
\end{array}
\right)
=\begin{pmatrix}
 \cos \alpha &0& -\sin \alpha \\ 0 &1& 0 \\ \sin \alpha &0& \cos \alpha
\end{pmatrix}
\left(
\begin{array}{c}
h_1 \\ h_2 \\ h_3
\end{array}
\right),
\end{equation}
In the new basis of Eq.(\ref{eq:higgs}), the Yukawa interaction terms in Eq.(\ref{eq:oyukawa}), can be re-expressed as
\begin{equation}
-\mathcal{L}_{Y}=\eta_{ij}^U\bar{\psi}^{i}_{L}\tilde{H}_1U^{j}_{R}+\xi_{ij}^U\bar{\psi}^{i}_{L}\tilde{H}_2U^{j}_{R}+\eta_{ij}^D\bar{\psi}^{i}_{L}H_1D^{j}_{R}+\xi_{ij}^D\bar{\psi}^{i}_{L}H_2D^{j}_{R}+\HC,
\end{equation}
where
\begin{eqnarray}
  \eta_{ij}^U&=& \eta_{ij}^{(1)}\cos\beta + \eta_{ij}^{(2)}e^{-i\delta}\sin\beta \equiv \sqrt{2}M_{ij}^U/v ,\nonumber \\
  \xi_{ij}^U &=&-\eta_{ij}^{(1)} \sin\beta+\eta_{ij}^{(2)}e^{-i\delta}\cos\beta,\nonumber \\
  \eta_{ij}^D&= & \xi_{ij}^{(1)}\cos\beta+\xi_{ij}^{(2)} e^{i\delta}\sin\beta \equiv \sqrt{2}M_{ij}^D/v, \nonumber\\
  \xi_{ij}^D &= &-\xi_{ij}^{(1)}\sin\beta+\xi_{ij}^{(2)}e^{i\delta}\cos\beta.
\end{eqnarray}
and $M^{U,D}$ are fermion mass matrices.
As the Yukawa coupling terms $\eta^{U}$ and $\xi^{D}$ become
complex due to the vacuum phase $\delta$ with real $\eta^{(1),(2)}$ and $\xi^{(1),(2)}$ defined in  Eq.~(\ref{eq:oyukawa}), the resulting mass matrices are also complex. Then unitary transformations are needed
to diagonalize the mass matrices,
\begin{eqnarray}
&& u_{L,R}^{j} = V_{L,R}^{jk,U}U_{L,R}^{k},\,\, V_{L}^{U}\eta^{U}V_{R}^{U\dagger} =\sqrt{2}\frac{m_{u}}{v}; \nonumber\\
&& d_{L,R}^{j} = V_{L,R}^{jk,D}D_{L,R}^{k},\,\, V_{L}^{D}\eta^{D}V_{R}^{D\dagger} =\sqrt{2}\frac{m_{d}}{v}.
\end{eqnarray}
By transforming electroweak interaction eigenstates of the fermions, $U_{L,R}$ and $D_{L,R}$, to their mass eigenstates, $u_{L,R}$ and $d_{L,R}$, we denote the final Yukawa couplings in the quark mass eigenstates by $\xi^{u,d}$, with the relation,
\begin{equation}
  \xi^{u,d} = V^{U,D}_L \xi^{U,D} V_R^{U,D\dag}
\end{equation}
With the above notation, the Yukawa interaction terms are
\begin{eqnarray}\label{eq:lagrangian}
 -\mathcal{L}_{Y}
 &=&\sum_{i=1}^{3}\left[m_u^{i}\bar{u}^{i}_{L}u_{R}^{i}(1+\frac{\phi_1^0}{v})+m_d^{i}\bar{d}^{i}_{L}d^{i}_{R}(1+\frac{\phi_1^0}{v})\right]\nonumber\\
 &&+\frac{1}{\sqrt{2}}\bar{u}^{i}_{L} {\xi}^u_{ij} u^{j}_{R}(\phi_2^0-i \phi_3^0)
   +\frac{1}{\sqrt{2}}\bar{d}^{i}_{L} {\xi}^d_{ij} d^{j}_{R}(\phi_2^0+i \phi_3^0)\nonumber\\
 &&- \bar{d}^{i}_{L}\hat{\xi}^u_{ij} u^{j}_{R}H^{-} +\bar{u}^{i}_{L}{\hat{\xi}^d_{ij}} d^{j}_{R}H^{+} + \HC.
\end{eqnarray}
where the charged Yukawa coupling are defined as
\begin{eqnarray}
\hat{\xi}^{u}=V_{\CKM}^{\dagger}\xi^{u};\, \, \hat{\xi}^{d} = V_{\CKM}\xi^{d}.
\end{eqnarray}
It can be seen that when there was no mixing among
$\phi^{0}_{1}$, $\phi^{0}_{2}$ and $\phi^{0}_{3}$, the scalar $\phi^{0}_{1}$ plays the role of the
Higgs in the SM.

In the following discussions, we shall use $\xi^{u},\xi^{d}$ and the quarks' masses as the free and independent input parameters instead of the original Yukawa coupling matrices ($\eta_{ij}^{(k)}$, $\xi_{ij}^{(k)}$) given in Eq.(\ref{eq:oyukawa}) and the parameter $\beta$. It is convenient to parameterize the Yukawa couplings $\xi_{ij}^{u,d}$ by using the quark mass scales\cite{Chang},
\begin{eqnarray}
\xi_{ij}^{u,d}\equiv \lambda_{ij} \sqrt{2m^{u,d}_i m^{u,d}_j }/v,
\end{eqnarray}
where the $i,j$ are the flavor indexes(for $\xi^u_{ij},i,j=u,c,t$ and for $\xi^d_{ij}, i,j=d,s,b$). And the smallness of the off-diagonal elements are characterized by the hierarchical mass scales of quarks and the parameters
$\lambda_{ij}$.

After the
transformation of the Higgs in Eq.(\ref{eq:transform}), the gauge
part of the Higgs in the basis of $\phi_i^0$ can be written as \cite{YLW1}
\begin{eqnarray}
  \mathcal{L}_{\rm G}&=&\left(D_\mu H_1\right)^\dag\left(D^\mu
  H_1\right)+\left(D_\mu H_2\right)^\dag\left(D^\mu
  H_2\right)\nonumber\\
  &=&\frac{1}{2}\partial_\mu\phi^0_1\partial^\mu\phi^0_1+\frac{(v+\phi^0_1)^2}{8}\left[(g^{\prime2}+g^2)Z^2+2g^2W^+W^-\right]\nonumber\\
  &+&\frac{1}{2}\left(\partial\phi_2^0\partial^\mu\phi^0_2+\partial\phi_3^0\partial^\mu\phi^0_3\right)+\partial_\mu H^-\partial^\mu H^+\nonumber\\
  &+&e^2H^+H^-A^2+\frac{(g^2-g^{\prime2})^2}{4(g^2+g^{\prime2})}H^+H^-Z^2+\frac{g^2}{2}H^+H^-W^+W^-+\frac{e(g^2-g^{\prime2})}{\sqrt{g^2+g^{\prime2}}}H^+H^-Z\cdot A\nonumber\\
  &+&\frac{g^2}{4}W^+W^-(\phi_2^{02}+\phi_3^{02}) + \frac{g^2+g^{\prime2}}{8}(\phi_2^{02}+\phi_3^{02})Z^2\nonumber\\
  &+&\left\{ieA^\mu H^-\partial_\mu H^++i\frac{g^2-g^{\prime2}}{2\sqrt{g^2+g^{\prime2}}}Z^\mu H^-\partial_\mu H^+ + \frac{ig}{2}W^-_\mu(\phi_2^0-i\phi_3^0)\partial^\mu H^+\right.\nonumber\\
  &+&\frac{eg}{2}A^\mu W^-_\mu H^+(\phi^0_2-i\phi_3^0) +\frac{g}{4}\frac{g^2-g^{\prime2}}{\sqrt{g^2+g^{\prime2}}}H^+W^-_\mu Z^\mu(\phi_2^0-i\phi_3^0)\nonumber\\
  &+&\frac{ig}{2}H^-W^+_\mu(\partial^\mu\phi_2^0+i\partial^\mu\phi_3^0)+\frac{i\sqrt{g^2+g^{\prime2}}}{4}(\phi^0_2-i\phi^0_3)Z^\mu(\partial_\mu\phi^0_2+i\partial_\mu\phi^0_3)\nonumber\\
  &-&\left.\frac{g\sqrt{g^2+g^{\prime2}}}{4}H^-W^+_\mu
  Z^\mu(\phi_2^0+i\phi_3^0)+\HC.\right\}.
\end{eqnarray}

\section{$W^\pm H^\mp$ associated production}
\begin{figure}[htb]
\includegraphics[scale=1]{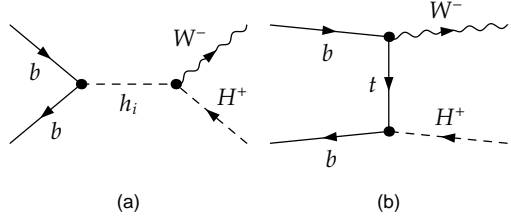}
\caption{Feynman diagrams for $W^{-}H^{+}$ production via $b\bar{b}$ annihilation at tree level, where $h_{i}$ denote $(h_1, h_2, h_3)$.}\label{Figbbwh}
\end{figure}

\begin{figure}[htb]
\includegraphics[scale=1]{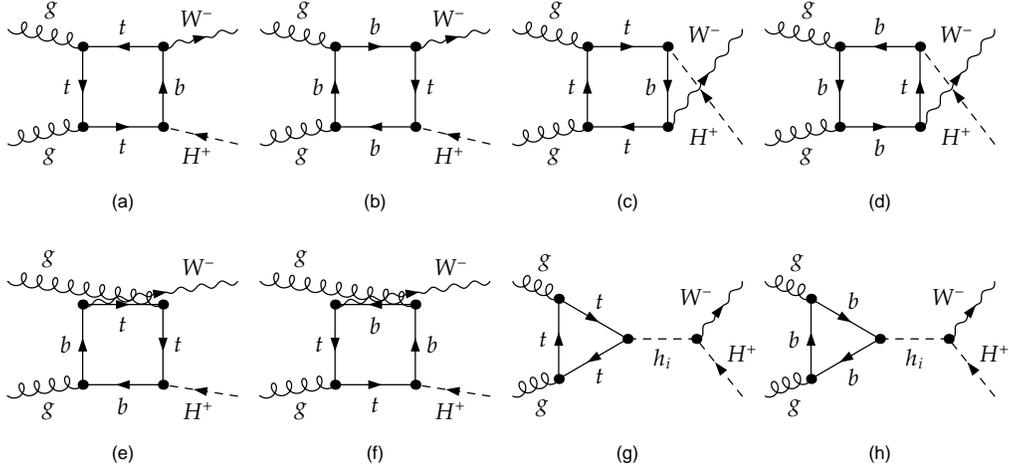}
\caption{Feynman diagrams for $W^{-}H^{+}$ production via $gg$ fusion at one-loop level.}\label{Figggwh}
\end{figure}

At hadron colliders, the dominant mechanisms for $W^\pm H^\mp$ associated production are $b\bar{b}$ annihilation at tree level and gluon-gluon fusion at one-loop level. As the Feynman diagrams shown in Fig.(\ref{Figbbwh}), the $b\bar{b}$ annihilation proceeds either via $s$-channel resonance mediated by the neutral Higgs $h_{i}$s, or by $t$-channel dominated by the top quark exchange. Here, we treat $b$ and $\bar{b}$ quarks as active partons inside the colliding protons and use the PDF(parton distribution functions) set cteq6m\cite{cteq6m}.

An alternative $W^\pm H^\mp$ production mechanism is provided
by gluon-gluon fusion shown in Fig.(\ref{Figggwh}). From the Feynman diagrams, although the parton-level cross section of gluon fusion is suppressed by $\alpha_s^2$ relative to
the one of $b\bar{b}$ annihilation, it is expected to yield a comparable
contribution at 14-TeV hadron colliders, due to the
overwhelming gluon luminosity. But as our result shown in Fig.\ref{FigbbMHp} and Fig.\ref{FigggMHp}, the $\sigma(gg\to H^\pm W^\mp)$ is much smaller than $\sigma(b\bar{b}\to H^\pm W^\mp)$, unless the Yukawa of top quark $\xi_t$ is very large.

The relevant interaction terms in this calculation are
\begin{eqnarray}
\mathcal{L}_{h_{j}H^{\pm}W^{\mp}}&=&
\frac{g}{2}\sum_{j}\left[ \tilde{g}_{j}(h_{j}i\overleftrightarrow{\partial_{\mu}}H^{-})W^{+,\mu}
-\tilde{g}_{j}^{*}(h_{j}i\overleftrightarrow{\partial_{\mu}}H^{+})W^{-,\mu}\right] \\
\tilde{g}_{j}&=&O_{2j}+iO_{3j}=g_{h_{j}H^{-}W^{+}} \nonumber
\end{eqnarray}
Yukawa terms(the light quark parts are ingored)
\begin{eqnarray}
\mathcal{L}_{h_{i}\bar{q}q}&=&-\frac{gm_{q}}{2m_{W}}\sum_{i}h_{i}\bar{q}\left[g_{i}^{q}P_{L}+g_{i}^{q*}P_{R}\right]q, \: q=b \textrm{ or } t \\
g_{i}^{b}&=&O_{1i}+\lambda_{b}^{*}(O_{2i}+iO_{3i}),\: \lambda_{b}\equiv \lambda_{bb} \nonumber \\
g_{i}^{t}&=&O_{1i}+\lambda_{t}^{*}(O_{2i}-iO_{3i}),\: \lambda_{t}\equiv \lambda_{tt} \nonumber
\end{eqnarray}
where  $P_{L}=(1-\gamma_{5})/2,\: P_{R}=(1+\gamma_{5})/2$, and
\begin{eqnarray}
\mathcal{L}_{H^{\mp}tb} = \frac{g}{\sqrt{2}m_{W}} H^{+}\bar{t}\left[\lambda_{t}^{*}m_{t}P_{L}-\lambda_{b}m_{b}P_{R}\right]b + \HC
\end{eqnarray}

As in the s-channel diagram of $b\bar{b}\to H^\pm W^\mp$, generally when we consider the decay width of Higgs particles in the Higgs propagators
 \begin{equation}
S_{h_{i}}=\frac{1}{p^2-M^2_i+iM_i\Gamma_{h_{i}}} =  \frac{p^2-M^2_i-iM_i\Gamma_{h_{i}}}{(p^2-M^2_i)^2+M_i^2\Gamma_{h_{i}}^2} ,\label{eq:propagator}
\end{equation}
which is the same for the $CP-$conjugate processes. Thus the resulting effective phase of complex production amplitude caused by the Higgs propagator with considering decay width will play the role of strong phase in the hadronic decays. This is different from the phase caused either from the mixing between $CP$-even and $CP$-odd Higgs states or from  the complex couplings of the electroweak Higgs eigenstates in Eq.(\ref{eq:lagrangian}) which has opposite sign between the $CP-$conjugate processes. In this case, there will be CP asymmetry in $W^\pm H^\mp$ productions such as considered in \cite{Akeroyd,Eriksson}. But at the LHC, as the central energy is very high, $s=(p_1+p_2)^2 \sim \TeV^2$, and the decay width of the Higgs is small, the CP asymmetry is suppressed by the factor $\Gamma^2_i/s$ which is hard to be detected. Since the CP violation comes from the interaction terms between different neutral Higgs contributions, to get large CP violation there must exist more than two heavy and very unstable neutral Higgs bosons.  On the other hand, as the $pp\to H^\pm W^\pm$ production is dominated by $b\bar{b}\to H^\pm W^\pm$, although the CP asymmetry in $gg\to H^\pm W^\pm$ is larger, the total CP asymmetry of $H^\pm W^\mp$ production on proton-proton collision remains small.

Furthermore, when the three Higgs bosons are all light, we can consider the first order in Eq.(\ref{eq:propagator}),
\begin{equation}
  S_{h_{i}}\sim \frac{1}{s}+\mathcal{O}(M^2_{i}/s).
\end{equation}
so that the three Higgs bosons have the similar propagators, which makes the effect of the mixing between $h_i$ and $h_j$ be very small, and suppressed by the factor $M^2_{i}M^2_{j}/s^2$ due to the orthogonality of the mixing matrix.

In this paper, we first take the neutral Higgs mixing matrix to be diagonal so as to see the effect of the CP phases of the Yukawa couplings which are absent in the Type-II 2HDM. Then we consider the effects of the mixing between the Higgs bosons and the dependence of the production on the Higgs masses.  In our calculations, the Feynman graphs are generated by using FeynArts \cite{Hahn:FeynArts} and evaluated with FormCalc and LoopTools \cite{Hahn:FormCalc}.

\section{Numerical Results}
It was observed in the Type III 2HDM \cite{YLW1,YLW2} that the
charged Higgs interactions involving the Yukawa couplings
$\hat{\xi}^{u,d}$ in Eq.(\ref{eq:lagrangian}) lead to a new type of
$CP$-violating FCNC even if the neutral current couplings
$\xi^{u,d}$ are diagonal. For the parameters concerning the third
generation, we may express as
\begin{eqnarray}
\xi_{tt}^u = \xi^t =\xi_{t}e^{\delta_t}, \quad \xi_{bb}^d = \xi^b =\xi_{b} e^{\delta_b}.
\end{eqnarray}
The general constraints on the FCNC and the relevant parameter
spaces have been investigated in \cite{PR,BCK,WUZHOU,2hdmeff,YLW4,b2gamma}. Here we may consider the
following three typical parameter spaces for the neutral Yukawa
couplings of $b$-quark and $t$-quark
${\xi^q}/{\sqrt{2}}=\lambda_{q}m_q/v$,
\begin{eqnarray}
\mbox{Case\, A}:\, |\xi^{t}/{\sqrt{2}}|&=& 0.2 (\lambda_t=0.3);\quad \ \,\quad |\xi^{b}/{\sqrt{2}}| =0.5 (\lambda_b=30),\nonumber\\
\mbox{Case\, B}:\, |\xi^{t}/{\sqrt{2}}|&=&0.1 (\lambda_t=0.15);\quad\quad |\xi^{b}/{\sqrt{2}}|=0.8 (\lambda_b=50), \label{eq:case}\\
\mbox{Case\, C}:\,
|\xi^{t}/{\sqrt{2}}|&=&0.01 (\lambda_t=0.015); \quad
|\xi^{b}/{\sqrt{2}}|=1.0 (\lambda_b=60),\nonumber
\end{eqnarray}
 which is consistent with the current experimental constraints in
the flavor sector including the $B$ meson decays\cite{B2PV,B2VV}
even when the neutral Higgs masses are light. In this note we take
\begin{eqnarray}\label{hmass1}
  m_{h_1}=115\GeV,\quad m_{h_2}=160\GeV,\quad m_{h_3}=120\GeV \mbox{ or } \, 500\GeV
\end{eqnarray}
as input. The strong coupling constant $\alpha_s(\mu)$ is running with $\alpha_s(M_Z)=0.1176$\cite{Yao:2006px}. And we identify the renormalization and factorization scales with the $W^\pm H^\mp$ invariant mass. And for the charged Higgs mass, the direct limit from LEP is $m_{H^\pm}>78.6\GeV$\cite{lep:2001ch} and can be as low as $100\GeV$ from B meson decay\cite{BCK}. The strong constraints may arise from the radiative bottom quark decay
$b\to s\gamma$. In fact, its mass was found to be severely constrained from the $b\to s\gamma$ decay in the Type II 2HDM, the lower bound on the charged Higgs mass can be as large as $m_{H^+} \simeq 350$ GeV, which is corresponding to the special case in the Type III 2HDM with the parameter $|\xi^t||\xi^b| \sim 0.02$(or $|\lambda_t\lambda_b|\sim1$) and a relative phase $\delta_t-\delta_b = 180^{\circ}$. In this note, we take the charged Higgs mass as free from 150GeV to 500GeV and take $m_{H^\pm}=200\GeV$ as a special case to study the $P_T$ distributions.

\begin{figure}[ht]
\includegraphics[scale=1]{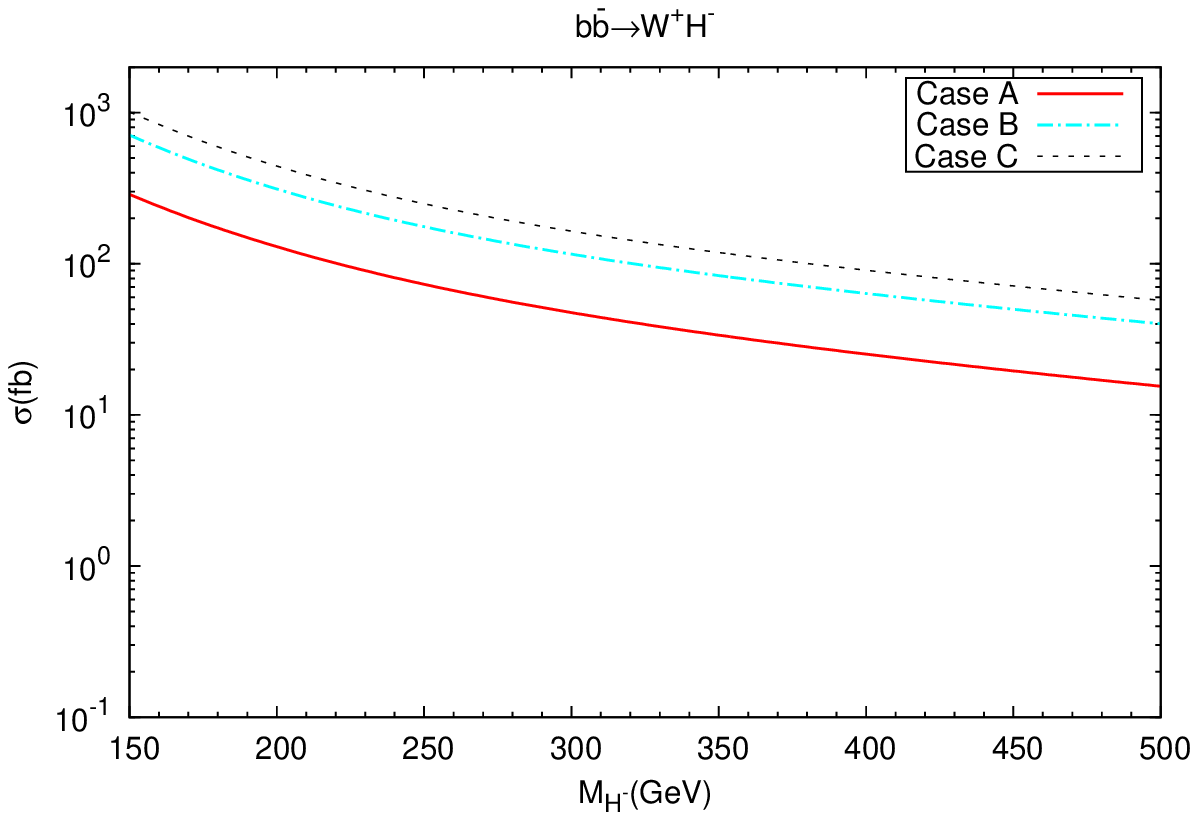}
\includegraphics[scale=1]{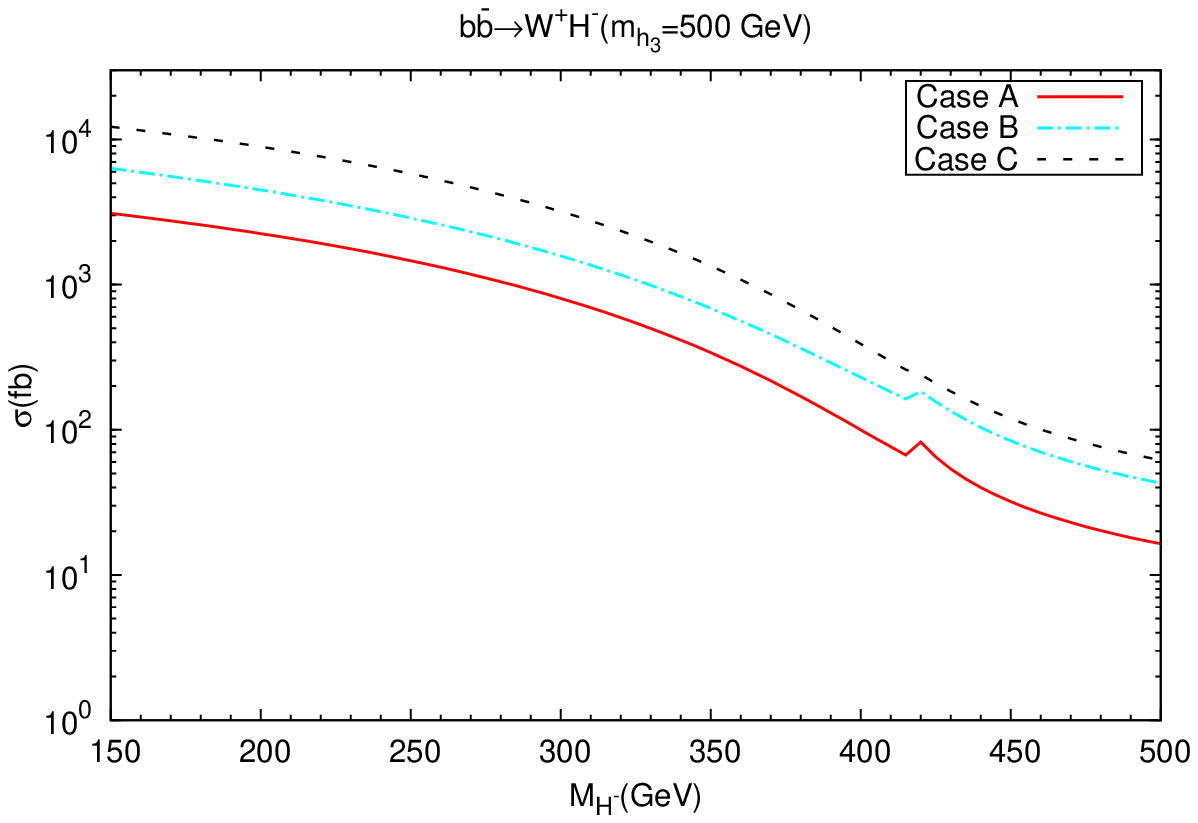}
\caption{The total cross section of $W^{+}H^{-}$ production from $b\bar{b}$ channel at the LHC ($\sqrt{S}=14 $ TeV) as function of the charged Higgs's mass in three cases, Case A(solid line), Case B(dashed-dot) and Case C(dashed) with $m_{h_3}=120\GeV,\, 500\GeV$.}
\label{FigbbMHp}
\end{figure}

\begin{figure}[ht]
\includegraphics[scale=1]{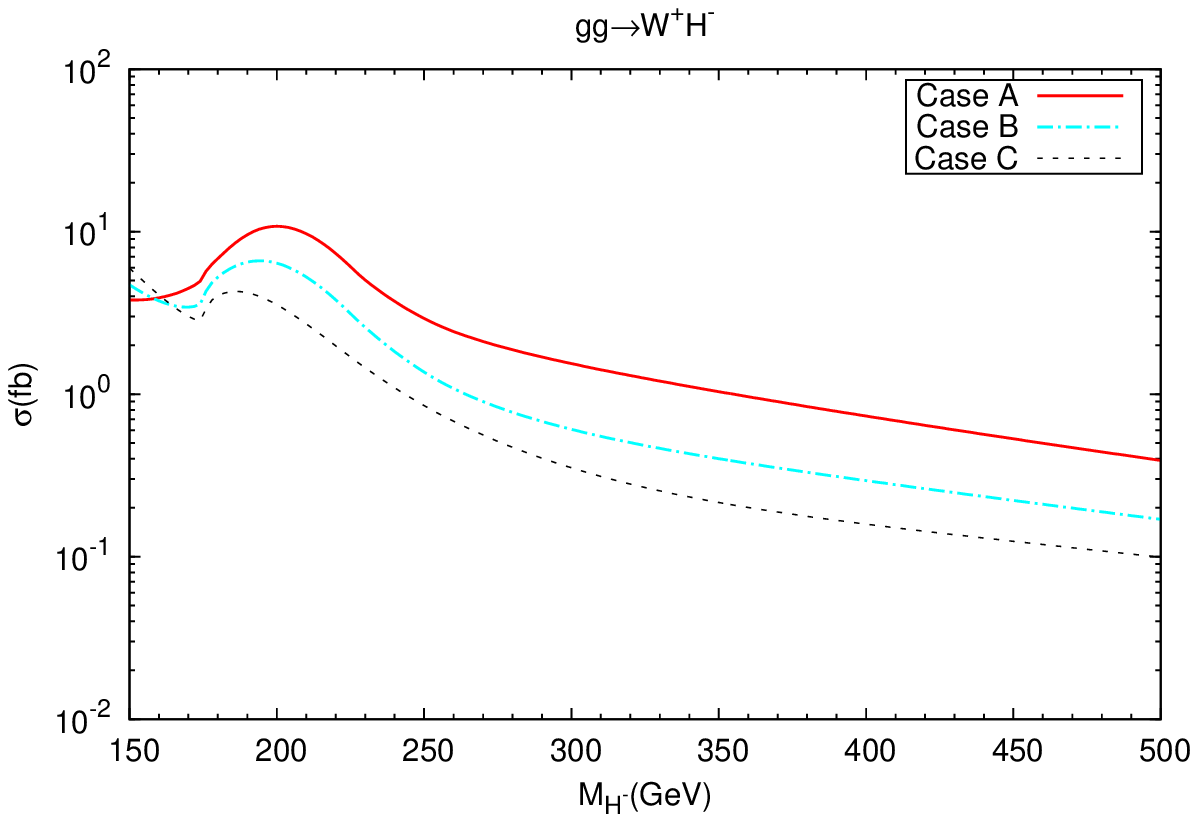}
\includegraphics[scale=1]{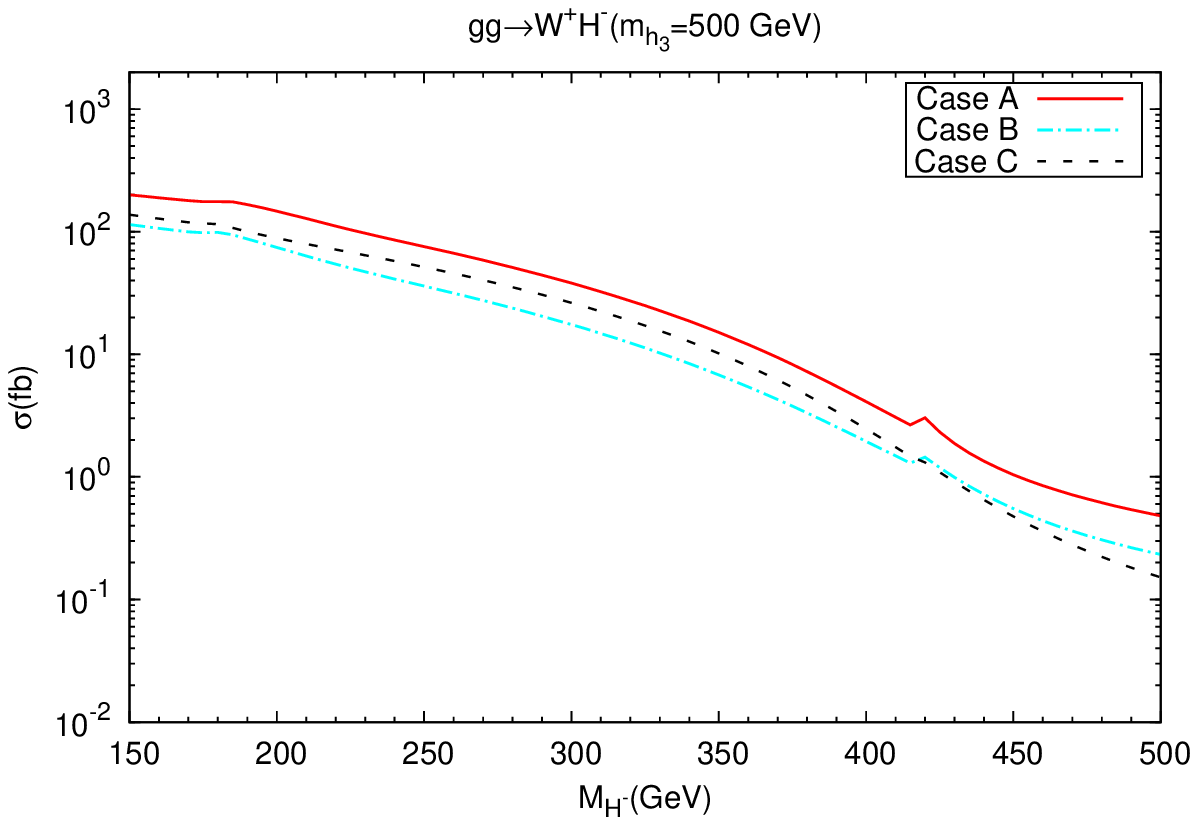}
\caption{The total cross section of $W^{+}H^{-}$ production from $gg$ channel at the LHC ($\sqrt{S}=14 $ TeV) as function of the charged Higgs's mass in three cases, Case A(solid line), Case B(dashed-dot) and Case C(dashed) with $m_{h_3}=120\GeV,\, 500\GeV$.}
\label{FigggMHp}
\end{figure}

In Fig.(\ref{FigbbMHp}) and Fig.(\ref{FigggMHp}), we show the dependence of the cross-section on the mass of the charged and neutral Higgs. Comparing the contribution from $b\bar{b}$ with $gg$, we can see that $b\bar{b}$-annihilation is the dominant channel and its contribution  is generally two order larger in Case B and C than that of $gg$-fusion. In Case A, it is about one order larger. In later discussion, we will only show the diagram from $b\bar{b}$ channel. The $b\bar{b}\to H^\pm W^\mp$ is dominant by the coupling $\xi_b$, therefore the $\sigma(b\bar{b}\to H^\pm W^\mp)$ of Case C is the largest and the Case A is the smallest. These results are also confirmed in Type-II 2HDM. As the $\tan\beta$ is smaller, the $\sigma(b\bar{b}\to H^\pm W^\mp)$ is smaller while the $\sigma(gg\to H^\pm W^\mp)$ is larger \cite{Kniehl}. Now we focus on the $b\bar{b}\to W^\mp H^\pm$.  We can see that when the neutral Higgs $h_3$ is heavier, the cross section is larger. That's because when $h_3$ is heavier, the propagator $S_{h_3}=1/(s-m^2_{h_3}+im_{h_3}\Gamma_{h_3})$ is larger and on-shell $h_3$ can be produced. The shape of the curves are also changed because of the effect of the width $\Gamma_{h_3}$ which play an important roles when $m_{h_3}=500\GeV$. At about $m_{H^\mp}\sim 420\GeV$ when $m_{h_3}=500\GeV$, there is a peak due to the threshold effect.

\begin{figure}[ht]
\includegraphics[scale=1]{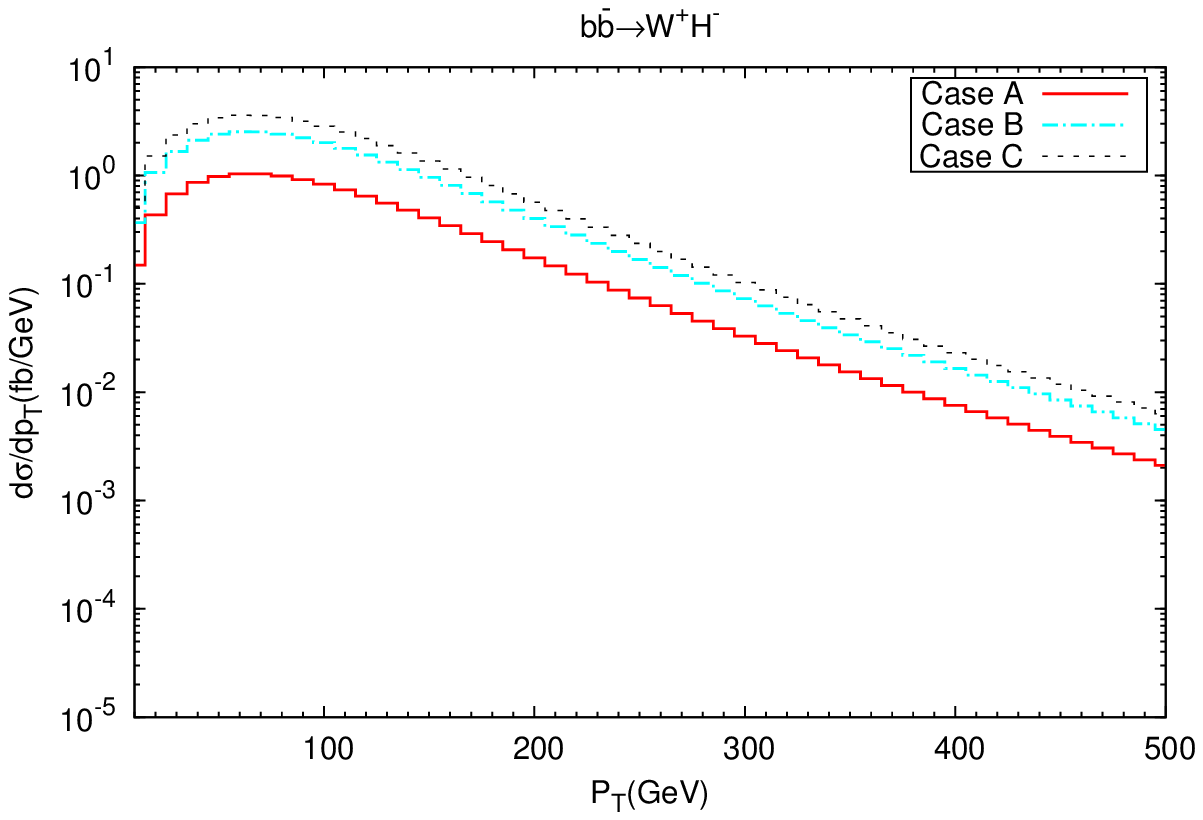}
\includegraphics[scale=1]{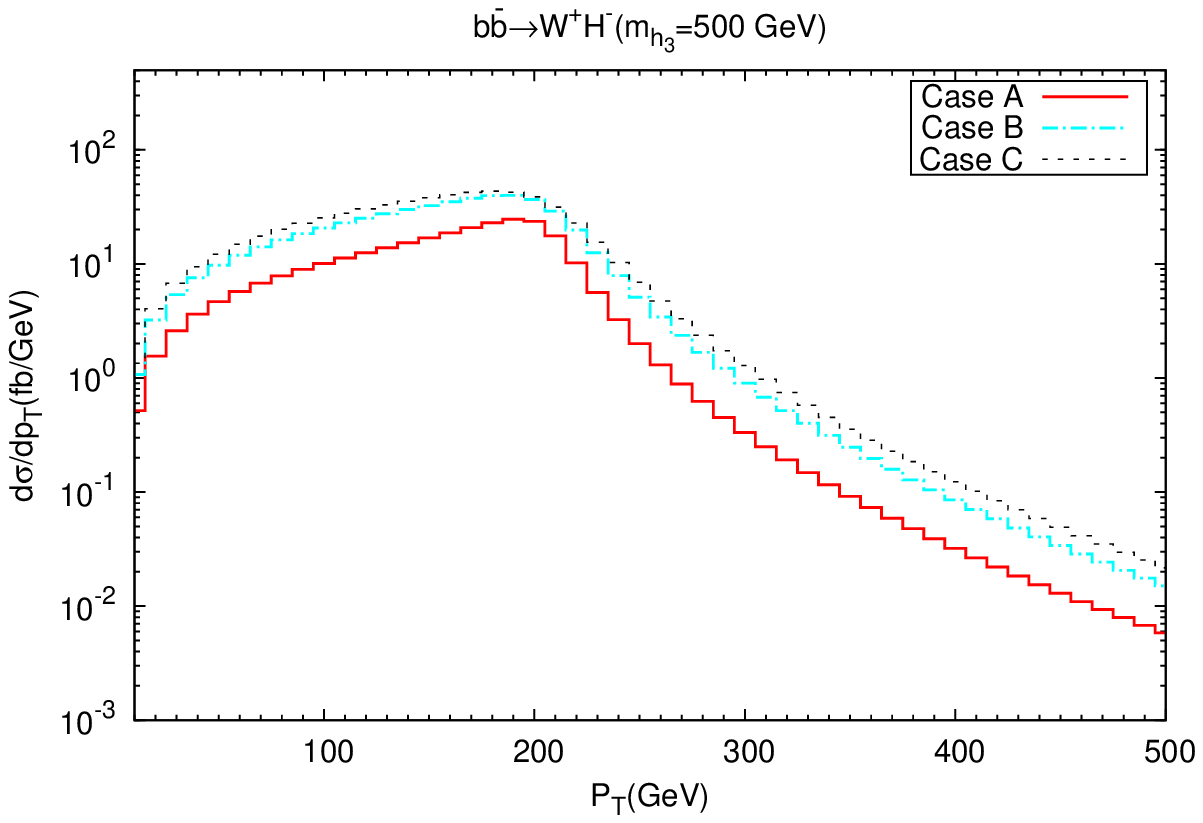}
\caption{The $P_T$ distribution of $W^{+}H^{-}$ production from $b\bar{b}$ channel at the LHC ($\sqrt{S}=14 $ TeV) with $m_{h_3}=120\GeV,\,500\GeV$ and fixed $m_{H^\pm}=200$GeV in three cases, Case A(solid), Case B(dashed-dot) and Case C(dashed).}
\label{FigbbPt}
\end{figure}

We also show the differential cross-section on the transverse momentum $p_T$ of $H^{-}$ in Fig.~(\ref{FigbbPt}) for $b\bar{b}$-channel. As the contribution from $gg$ fusion is small, we do not consider its $P_T$ distribution here and the final result of $pp\to W^\pm H^\mp$ is dominated by $b\bar{b}$ contribution. The curves have different shapes in the two cases, $m_{h_3}=500\GeV$ and $m_{h_3}=120\GeV$.

As in general the $\lambda_{ij}$ can be complex, we shall consider the dependence of the cross-section on the phase difference between $\lambda_{bb}$ and $\lambda_{tt}$($\delta=\delta_b-\delta_t$). Generally we can take $\delta_b=\delta$ and $\delta_t=0$. If $\delta_t\neq0$, the curve will be globally shifted, and its shape will not be changed. The cross section of $b\bar{b}\to H^\pm W^\mp$ channel varies less than $1\%$ as $\delta \in [0,2\pi]$, as the $s$- and $t$-channel are all dominant by $\xi_{b}$, and the cross terms are suppressed as $O(\frac{m_b}{m_t})$. Since the total cross section from $gg$ is much small than that from $b\bar{b}$ channel in Case B and Case C, $\lambda_{bb}$ phase has almost negligible effect on the total production of $W^{\mp}H^{\pm}$.

\begin{figure}[ht]
\includegraphics[scale=0.75]{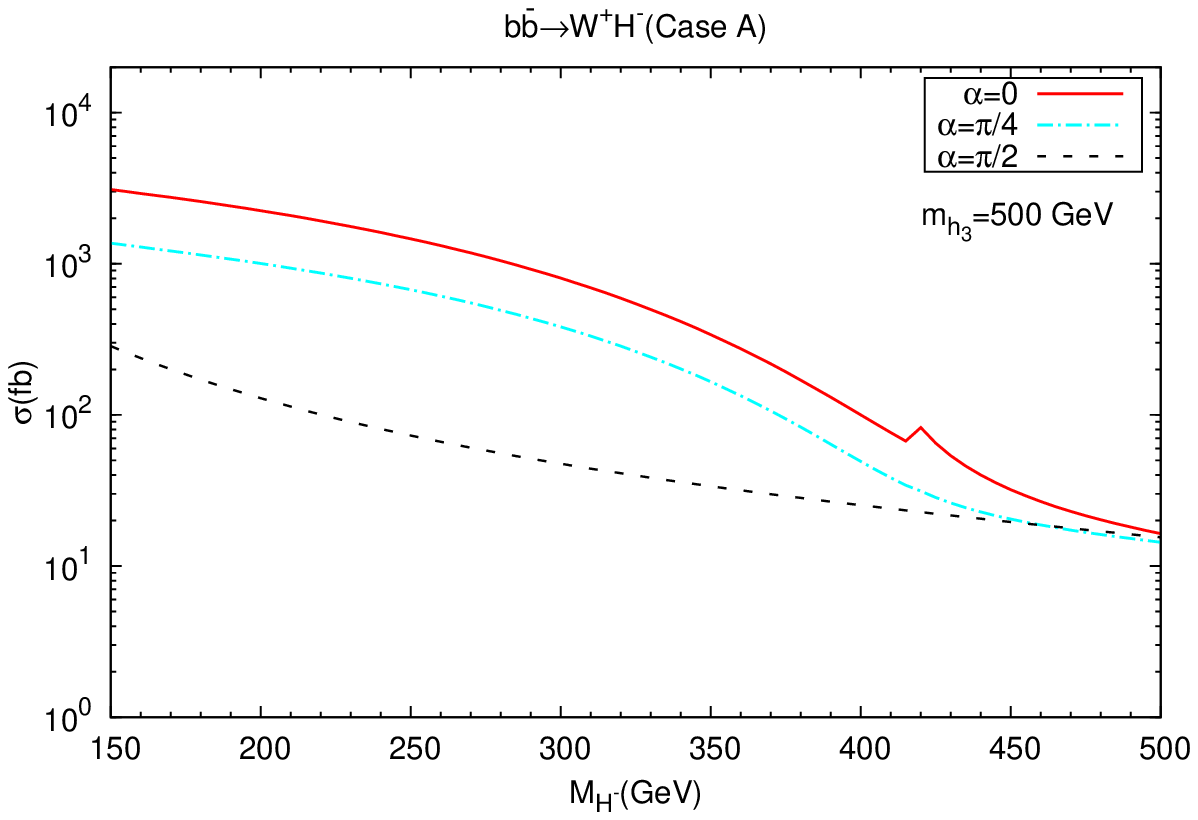}
\includegraphics[scale=0.75]{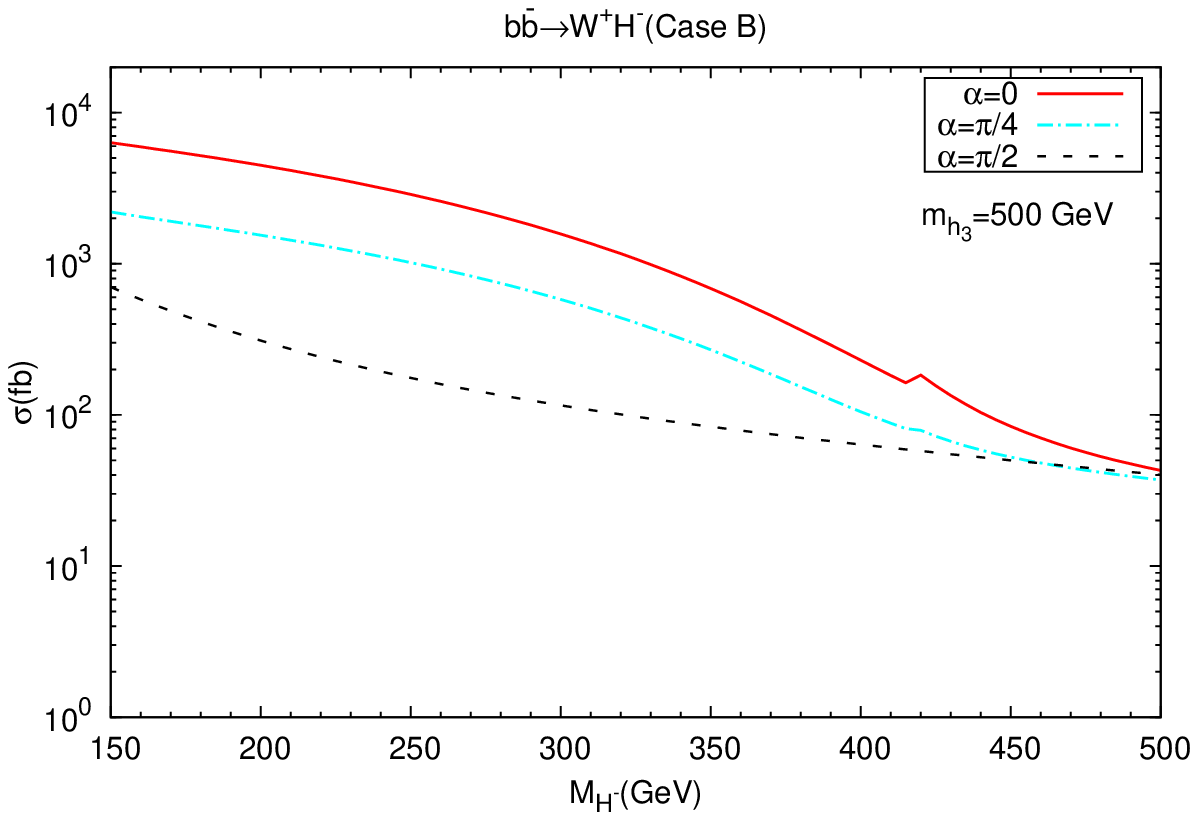}
\includegraphics[scale=0.75]{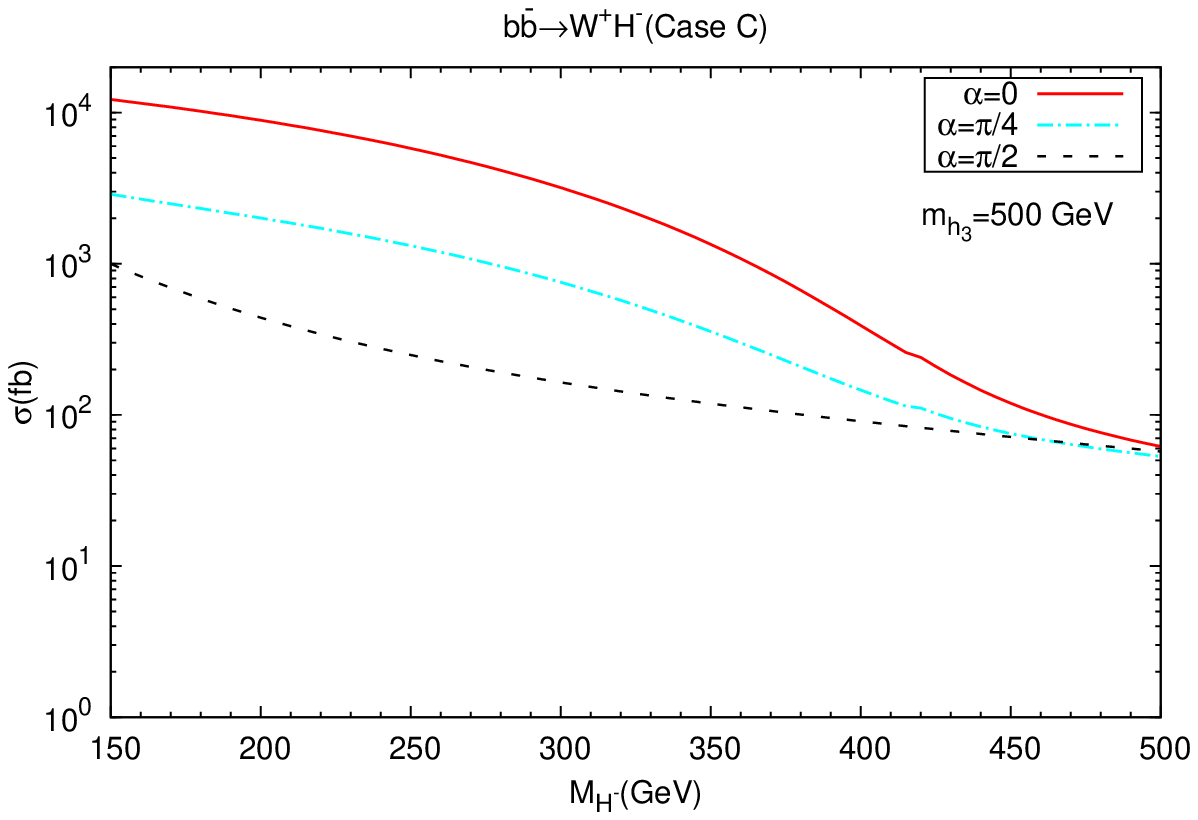}
\caption{The total cross section of $W^{+}H^{-}$ production from $b\bar{b}$ channel at the LHC ($\sqrt{S}=14 $ TeV) as the charged Higgs's mass and mixing angle ($\alpha$) between $h_1$ and $h_3$ in three cases, $\alpha=0$(solid line), $\alpha=\frac{\pi}{4}$(dashed-dot) and $\alpha=\frac{\pi}{2}$(dashed).}
\label{Fig500bbMHp}
\end{figure}

Now we would like to discuss the effect of the mixing between neutral higgs bosons. As discussed in last section, the $h_1$ and $h_2$ are light, their mixing effect can be neglected. Therefore, for $m_{h_3}=120\GeV$, the mixing effect to $pp\to W^\mp H^\mp$ can not be detected. When $m_{h_3}=500\GeV$, the mixing of $h_1$ or $h_2$ with $h_3$ can be sizable. Here we shall consider the mixing between $h_1$ and $h_3$ as it can be seen from the lagrangian that without mixing, $h_1$ has no contribution to $W^\mp H^\mp$ production. The width effect of higgs bosons will also be included. The total decay widths of $h_3$ are listed in Table 1 for different cases and mixing angles $\alpha$ between $h_1$ and $h_3$. For simplicity, we only consider the dominant decay channel at tree level $h_3 \rightarrow b\bar{b}$, $h_3 \rightarrow t\bar{t}$, $h_3 \rightarrow WW^{(*)}$, $h_3 \rightarrow ZZ^{(*)}$ and $h_3 \rightarrow W^\mp H^\pm$. The $W^{(*)}$ and $Z^{(*)}$ means that the bosons can be on-shell or off-shell as treated in \cite{Djouadi}. As the $b\bar{b}$, $t\bar{t}$, $WW^{(*)}$ and the $ZZ^{(*)}$ modes are not dependent on $m_{H^\pm}$, we sum them together and list with three cases.  While the decay mode $h_3 \rightarrow W^\mp H^\pm$ is dependent on the $m_{H^{\mp}}$ and independent on the different cases, thus we list it separately in the last column only when $m_{H^{\mp}}=200\GeV$.
\begin{table}[htb]
\begin{center}
 \begin{tabular}{|l|c|c|c|c|}
 \hline\hline
 $\Gamma _{h_3}(\GeV)$ &$\:$ Case A $\:$& $\:$ Case B $\:$ &$\:$ Case C $\:$ &$W^\mp H^\pm(200\GeV)$\\ \hline
 $\alpha =0 $          & 17.552 & 43.961 & 62.626 & 44.094 \\ \hline
 $\alpha =\pi/4$       & 41.764 & 54.969 & 64.301 & 22.047 \\ \hline
 $\alpha =\pi/2$       & 65.135 & 65.135 & 65.135 & 0 \\ \hline
 \end{tabular}
 \caption{The decay width of the Higgs $h_3$ when $m_{h_3}=500\GeV$. We list the fractional width $W^\pm H^\mp$ and the sum width of $b\bar{b}$, $t\bar{t}$, $WW^{(*)}$ and $ZZ^{(*)}$ separately.}\label{A0Width}
\end{center}
\end{table}
With the above decay width, we present our results in Fig.~(\ref{Fig500bbMHp}) for $b\bar{b}$ channel with $m_{h_3}=500\GeV$. It can be seen that the effect of the mixing for the three cases is similar, as the mixing angle $\alpha$ goes larger, the contribution from the $h_3$ gets smaller and the total cross section goes down. And when $\alpha=\pi/2$, the dashed lines are almost the same as the three lines shown in Fig.(\ref{FigbbMHp}) and Fig.(\ref{FigggMHp}) with $m_{h_3}=120\GeV$. This is because in Fig.(\ref{FigbbMHp}) and Fig.(\ref{FigggMHp}), there are only $h_2$ and $h_3$ ($m_{h_3}=120\GeV$) contribute to the production, and in Fig.(\ref{Fig500bbMHp}) with $\alpha=\pi/2$, the roles played by $h_3$ and $h_1$ are just interchanged, and the numerical results are also almost the same as $m_{h_1}=115\GeV\sim 120\GeV$.

\section{Conclusion}
In this paper, we have studied the production of a charged Higgs boson in association with a $W$ boson at the LHC in the Type-III 2HDM with spontaneous CP violation. We find that the cross sections of $W^{\mp}H^{\pm}$ production are large enough to consider opportunity to observe these processes, and also find no observable effects of CP violation in these processes. In this model, the charged Higgs boson mass can be as low as about 150 GeV due to the effective complex Yukawa couplings, which distinguishes from the Type II 2HDM with charged Higgs mass being constrained to be larger than 350 GeV via rare B meson decays. The possibility of detecting charged Higgs have been studied extensively in Type II 2HDM or MSSM with leptonic decay $H^+\to \tau^+\nu$ or hadronic decay $H^+\to t\bar{b}$ \cite{Barger:1993th, Belyaev:2002eq,Assamagan:2004tt,Abazov:2008rn, Raychaudhuri:1995cc, Moretti:1996ra, Roy:1999xw, Guchait:2006jp, Roy:1999xw, Eriksson,Gao:2007wz}. As our main interest is the effect of the general Yukawa couplings with the spontaneous CP-violation, we have considered only at the parton level, and will leave the inclusion of parton showering, hadronization, full simulation of the detector, etc. for future investigation.
As the new physics inputs have large uncertainties in the parameter space, we have not tried to include any higher order corrections in the production cross-section, and only included the contributions from $b\bar{b}$ annihilation and gg fusion to the lowest order. Using up-to-date information on the input parameters and
proton PDFs, we have presented theoretical predictions for the $W^\pm H^\mp$ production cross section. It has been found that unless very large top-Yukawa coupling, the $gg$ fusion has small contributions and the $b\bar{b}$ annihilation is the dominated mechanism for $W^\mp H^\pm$ production at LHC. Apart from the fully integrated cross section, we have also analyzed distributions in $p_T$ and considered the effect of the mixing between light CP-even $h_1$ and CP-odd $h_3$. As a consequence, it has been shown that the mixing effect is general small unless the mass gap of the neutral Higgs becomes large.

\begin{acknowledgments}
The authors would like to thank Ci Zhuang for useful
discussions. Y. Tang would like to thank Y.Q. Ma for helpful discussions. This work was supported in part by the National Science Foundation of China (NSFC) under the grant \# 10821504, 10975170 and the key Project of
Knowledge Innovation Program (PKIP) of Chinese Academy of Science.
\end{acknowledgments}

\newpage

\end{document}